\documentclass[aps,prb,twocolumn,floatfix,showpacs,groupedaddress]{revtex4}
\usepackage[dvips]{graphicx}

\begin{document}
\title{End group effect on electrical transport through individual 
molecules: A microscopic study}

\author{Yongqiang Xue $^{*}$ and Mark A. Ratner}
\affiliation{Department of Chemistry and Materials Research Center, 
Northwestern University, Evanston, IL 60208}

\begin{abstract}
The effect on molecular transport due to chemical modification of the 
metal-molecule interface is investigated, using as an example the prototypical 
molecular device formed by attaching a p-disubstituted benzene molecule 
onto two gold electrodes through chemically different end groups. Using a 
first-principles based self-consistent matrix Green's function method, we 
find that depending on the end group, transport through the molecule 
can be mediated by either near-resonant-tunneling or 
off-resonant-tunneling and the conductance of the molecule varies over 
more than two orders of magnitude. Despite the symmetric device 
structure of all the molecules studied, the applied bias voltage can 
be dropped either equally between the two metal-molecule contacts or 
mostly across the source (electron-injecting) contact depending on the 
potential landscape across the molecular junction at equilibrium.   
\end{abstract}
\date{\today}
\maketitle


The continuing development of molecular electronics represents the 
convergence of the trend of device miniaturization and the growing 
expertise on single-molecule manipulation through scanning nanoprobes, 
mechanical break junctions and supramolecular self-assembly 
techniques.~\cite{Reed99} In contrast to carbon nanotubes and other 
inorganic semiconductor nanowires 
with uniform lattice structure,~\cite{Dekker99} devices based on 
individual organic molecules often require attaching appropriate end groups 
chemically different from the molecule core in order to establish 
stable contact to the metallic electrodes.~\cite{Reed99}  
The introduction of end groups into the molecular structure has two 
immediate consequences: 
(1) it introduces molecular states that are end group-based; (2) it 
modifies the metal-molecule interaction through the metal-end group bond.   
Single-molecule devices can therefore be considered 
as atomic-scale heterostructured devices, ~\cite{Xue02,Xue031} 
where the heterostructure can be introduced at the metal-molecule interface 
through the end groups or in the molecule core through appropriate 
molecular design techniques. The purpose of this paper is thus to 
elucidate the prospect of interface ``engineering'' of molecular transport 
due to chemical modification of the metal-molecule interface,  
through detailed microscopic study of selected single-molecule devices. 

\begin{figure}
\includegraphics[height=3.0in,width=3.0in]{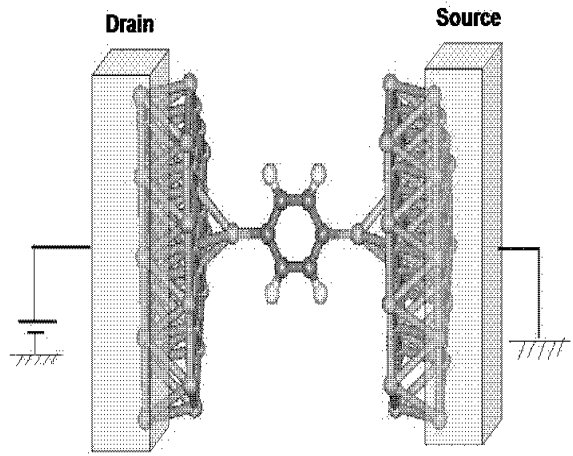} 
\caption{\label{xueFig1} (Color online) Schematic 
illustration of the metal-molecule-metal junction. Six gold 
atoms on each metal surface are included into the ``extended molecule'' 
where the self-consistent calculation is performed. The molecule can be 
attached onto the electrodes through chemically-different end groups.}
\end{figure}

The interface ``engineering'' considered here is achieved through modifying 
the valence structure of the end group. The devices we consider 
are formed by attaching the p-disubstituted benzene molecule onto two 
semi-infinite gold electrodes through oxygen (O) and fluorine (F) end 
atoms and the isocyanide (C-N) end group. The device structure 
is shown schematically in Fig.\ \ref{xueFig1}. In the case 
of C-N end group, the end atom can be either carbon (C) or nitrogen 
(N). Since the hydroxyl substituent (O-H) may not deprotonate on 
contact with the metallic electrodes, we consider both O end atom 
and O-H end group in this work. For clarity in notation, we will denote the 
five molecules following their structure as O$\Phi$O, HO$\Phi$OH, F$\Phi$F, 
CN$\Phi$NC and NC$\Phi$CN ($\Phi$ stands for benzene ring) respectively. 
The molecules chosen are thus among the smallest possible where 
metal-molecule interaction should have a strong effect on electrical transport, 
but are still representative of current experimental efforts.~\cite{Reed99}

As the end group changes, both the energy and the 
charge distribution (symmetry) of the frontier molecular orbitals (occupied 
and unoccupied molecular orbitals energetically closest to the metal 
Fermi-level, $-5.31$ eV for gold$\langle 111 \rangle$ electrode) change, 
depending on the chemical difference between the end group and the core 
atoms. For the NC$\Phi$CN and CN$\Phi$NC molecules, both the 
highest-occupied-molecular-orbital (HOMO) and 
lowest-unoccupied-molecular-orbital (LUMO) are delocalized through the entire 
molecule (central benzene as well as end group) and involve mainly 
the $P_{\pi}$ orbitals of carbon and nitrogen. Switching to O and F atoms, 
the HOMOs remain delocalized through the entire molecule, but the LUMOs 
become localized within the molecular core (central benzene). 
The HOMO-LUMO gap increases as we move from O to F (this is expected since these 
electron-withdrawing substituents stabilize the occupied levels). The frontier 
molecular states of the O$\Phi$O molecule (spin-singlet) show similar features 
to the phenyl dithiolate (PDT or S$\Phi$S in the present notation) 
molecule~\cite{Xue02,Xue031} due to the 
identical valence structure of the O and S atoms. Adding the H end atom 
increases the HOMO-LUMO gap of the HO$\Phi$OH molecule. Although the HOMO 
of the HO$\Phi$OH molecule remains delocalized 
over the oxygen and carbon atoms, it has little weight on the H ends which 
effectively reduces its coupling with the metal surface states.      

Due to differences in the energy and charge distribution of the frontier 
molecular states, the above molecules show distinctly different behavior in the 
molecular response to the perturbation induced by the metal-molecule 
interaction and the applied bias voltage. This is investigated using a 
first-principles based  self-consistent matrix Green's function theory which 
combines the Non-Equilibrium Green's Function theory of quantum transport 
with an effective single-particle description of molecular junction electronic 
structure using density-functional theory (DFT).~\cite{Xue02,Xue031}  
The theoretical models used have been discussed extensively in our previous 
work,~\cite{Xue02,Xue031} where we presented the general 
theoretical approach for modeling transport through single-molecule 
devices within the coherent transport regime and identified two key device 
processes for understanding the transport characteristics of a 
two-terminal molecular device: the equilibrium energy-level lineup and 
the nonequilibrium charge/potential response to the applied bias. The 
same theoretical and modeling approaches are adopted here. To summarize, 
the gold electrodes are modeled as semi-infinite $\langle 111 \rangle$ 
single-crystals. Six nearest-neighbor gold atoms on each metal surface (twelve 
gold atoms overall) are included into the ``extended molecule'' where the 
self-consistent calculation is performed. The rest of the electrodes (with the six 
atoms on each surface removed) are considered as infinite electron reservoirs, 
whose effects are included as self-energy operators. The calculation is 
performed using a modified version of Gaussian98~\cite{G98} 
using the Becke-Perdew-Wang parameterization of density-functional 
theory ~\cite{BPW91} and appropriate pseudopotentials with 
corresponding optimized Gaussian basis sets.~\cite{Xue031,Note1} 
For comparison with our previous work and other experimental and theoretical 
works,~\cite{Reed99,Theory} we keep the same adsorption geometry 
and metal surface-end group distance for all the molecules 
studied.~\cite{Xue031,Note2} The calculation is performed at 
room temperature. 

\begin{figure}
\includegraphics[height=3.0in,width=3.0in]{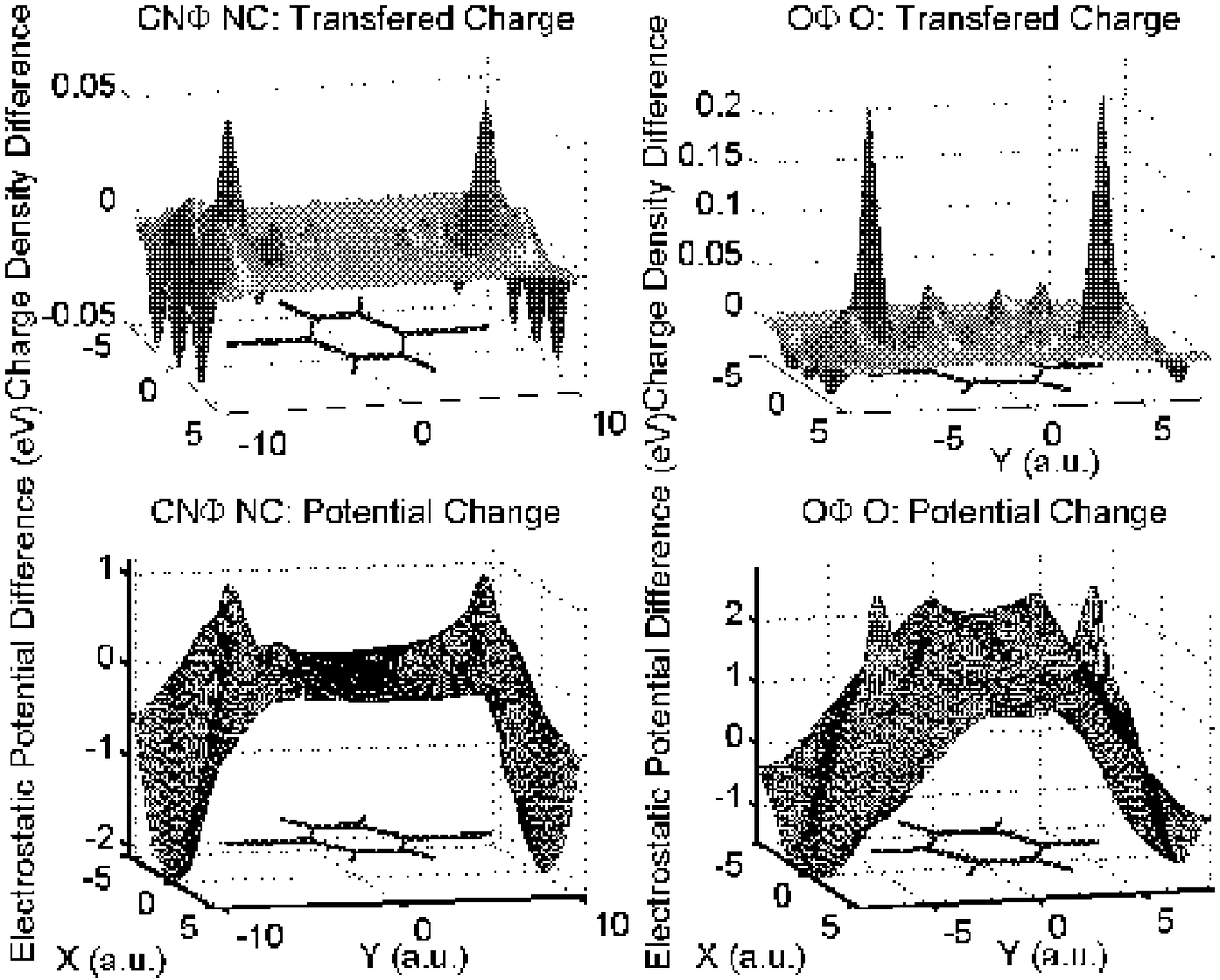}
\caption{\label{xueFig2} (Color online) Charge transfer (in unit of 1/(a.u.)$^{2}$) 
and cross-sectional view of the electrostatic potential change upon the formation of the 
gold-CN$\Phi$NC-gold (left figure) and gold-O$\Phi$O-gold (right figure) junctions. 
Here the spatial distribution 
of the transferred electrons is plotted as a function of position in the 
xy-plane (defined by the benzene ring) after integrating over the z-axis. 
Also shown is the projection of the molecule onto the xy-plane.}
\end{figure}

\begin{figure}
\includegraphics[height=3.0in,width=3.0in]{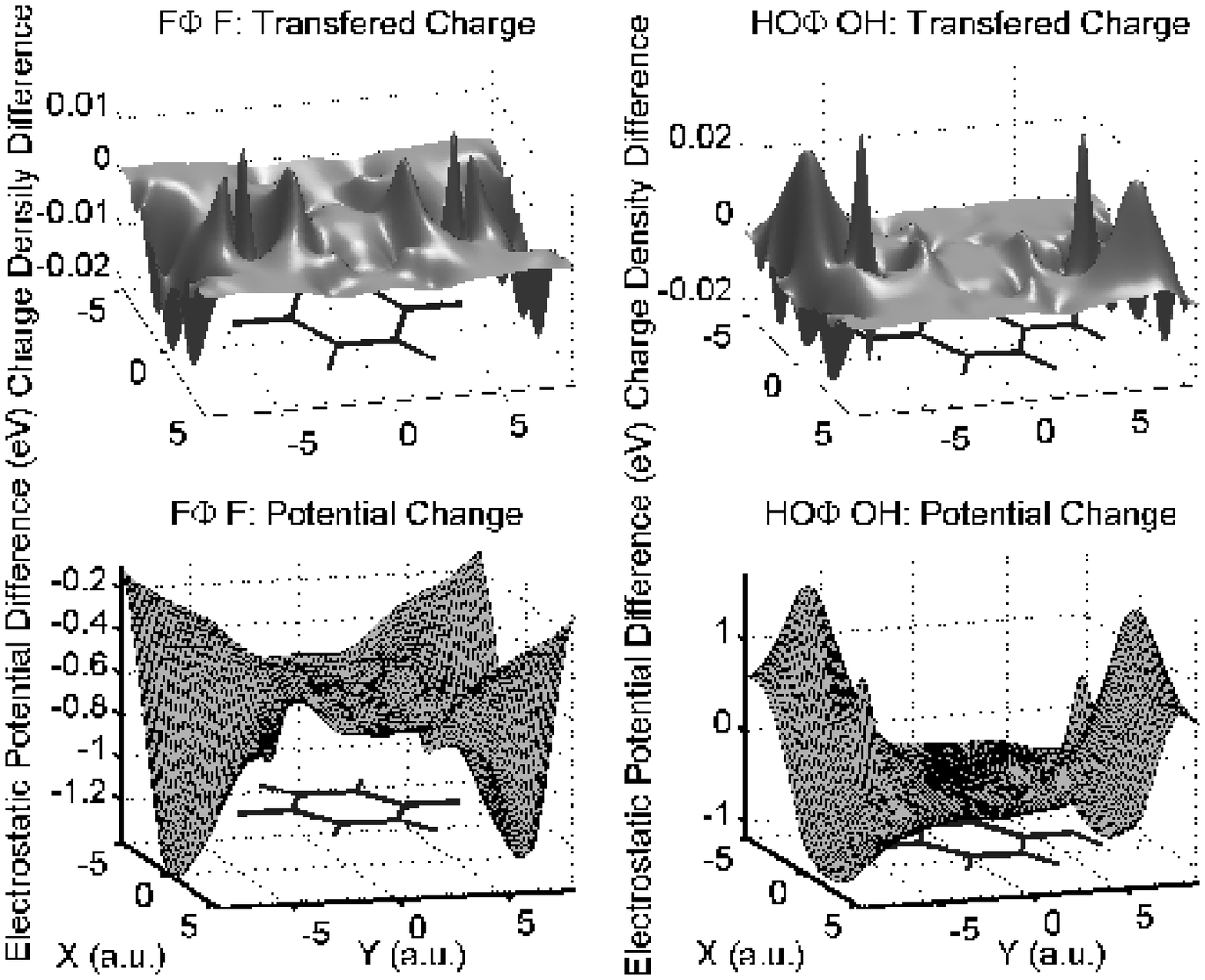}
\caption{\label{xueFig3} (Color online) Charge transfer (in unit of 1/(a.u.)$^{2}$) 
and cross-sectional view of the electrostatic potential change 
upon the formation of the gold-F$\Phi$F-gold (left figure) and 
gold-HO$\Phi$OH-gold (right figure) junctions. Here the spatial distribution 
of the transferred electrons is plotted as a function of position in the 
xy-plane (defined by the benzene ring) after integrating over the z-axis. 
Also shown is the projection of the molecule onto the xy-plane.}
\end{figure}

Following Ref.\ \onlinecite{Xue031}, we analyze the device physics of 
the metal-molecule-metal junction both at equilibrium and out of equilibrium.    
The most important quantity at equilibrium is the charge 
transfer between the molecule and the electrodes upon formation of the 
metal-molecule-metal junction, which may be understood qualitatively 
from the bonding configuration change across the metal-molecule interface 
and analyzed through the symmetry and energy of the frontier molecular 
orbitals. The transferred charge and the induced change in the electrostatic 
potential are obtained by taking the difference between the self-consistent 
charge/potential distribution in the equilibrium metal-molecule-metal 
junction and the charge/potential distribution in the isolated molecule plus 
the bare bimetallic junction.  For the CN$\Phi$NC and NC$\Phi$CN molecules, 
electron densities increase on the the end atom $P_{\pi}$ orbitals 
(Fig.\ \ref{xueFig2}). For the O$\Phi$O molecule, the overall spatial 
distribution of the transferred charge is similar to that of the PDT molecule 
(with electron density increasing on the end atom $P_{\pi}$ orbitals but 
decreasing in the $P_{x}$ orbitals),~\cite{Xue031} but the magnitude of 
the charge transfer is larger due to the larger electronegativity of the 
oxygen atom which leads to a larger barrier for electron injection from the 
metal into the molecule (Fig.\ \ref{xueFig2}). 
Keeping the H atom reduces significantly the amount of charge transferred to the 
HO$\Phi$OH molecule due to the saturated H-O bond (Fig.\ \ref{xueFig3}). Since 
most of the charge transfer is to the end H and O atoms, this leads to a large 
potential barrier at the H-O bonding region (Fig.\ \ref{xueFig3}).  The F$\Phi$F 
molecule shows different behavior compared to the other molecules. The lowest 
unoccupied state is core-based, and is located about $3.8(eV)$ above the 
metal Fermi-level. So the bonding configuration change is mainly realized by 
the charge redistribution among the occupied molecular states, leading to 
an electron density decrease in the fluorine-carbon $\sigma$ bond and an 
electron density increase in the fluorine-surface bond. The magnitude 
of the charge transfer is much smaller than the O$\Phi$O molecule 
and gives rise to an electrostatic potential well instead of electrostatic 
potential barrier for electron injection into the molecule (Fig.\ \ref{xueFig3}). 

\begin{figure}
\includegraphics[height=3.0in,width=3.0in]{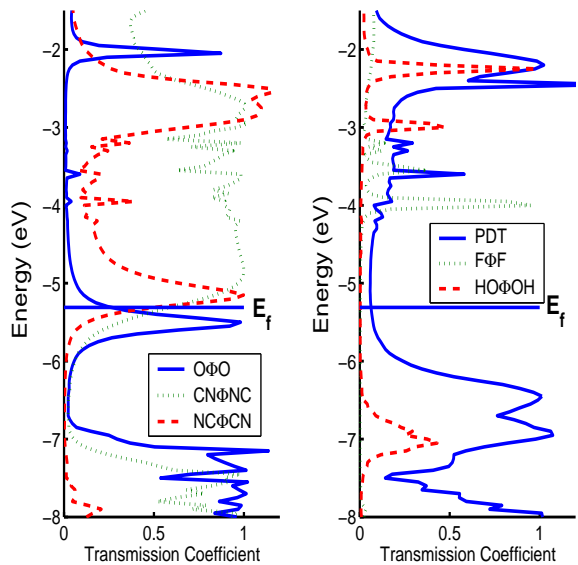}
\caption{\label{xueFig4} Transmission versus energy characteristics 
of the equilibrium molecular junction (at zero bias). The horizontal lines 
show the location of the metal Fermi-level.}
\end{figure}

The charge transfer processes and the resulting change in the 
electrostatic potential determine the energy-level lineup relative to the 
metal Fermi-level for the equilibrium molecular junction as well as low-bias 
conductance, which is obtained from the electron transmission 
characteristics at zero bias and calculated using the method of 
Refs.\ \onlinecite{Xue02} and \onlinecite{Xue031}. The results 
are illustrated in Fig.\ \ref{xueFig4}. For comparison we have also shown 
the transmission characteristics of the PDT molecule. 
For both the CN$\Phi$NC and NC$\Phi$CN molecules, 
the metal Fermi-level $E_{F}$ is in near-resonance with the LUMO state,  
which is delocalized over the entire molecule. This leads to broad transmission 
spectrum around $E_{F}$ and large zero-bias conductance (close to the 
conductance quantum) due to near-resonant-tunneling 
through LUMO. The difference in the transmission spectrum above $E_{F}$ 
between the two molecules is due to the difference in the energy-level 
spacing and broadening of the unoccupied states (the unoccupied states 
of CN$\Phi$CN are more closely spaced and more broadened than those 
of NC$\Phi$CN). The HOMO-LUMO gap of 
NC$\Phi$CN is also larger than that of CN$\Phi$CN, pushing the HOMO-mediated 
transmission peak further below $E_{F}$. For the O$\Phi$O and F$\Phi$F 
molecules where the end group atoms are more electron-rich, transmission at the 
metal Fermi-level is due to tunneling through the HOMO-LUMO gap.  
Although transmission through the middle of the gap is small, the HOMO 
of the O$\Phi$O molecule lines up much closer to the metal Fermi-level than PDT, 
leading to a much larger conductance. Keeping the end H atoms (HO$\Phi$OH) 
pushes both peaks corresponding to HOMO and LUMO down, with $E_{F}$ lying 
slightly closer to HOMO than LUMO, which reduces the conductance of the 
HO$\Phi$OH junction by a factor of $\approx 100$. Finally for the F$\Phi$F 
molecule where the end atom has the largest number of valence electrons, 
the LUMO instead lines up closer to the metal Fermi-level than the HOMO. 
Since the LUMO state of F$\Phi$F is localized on the molecule core and the 
HOMO state is pushed far below the metal Fermi-level (beyond the energy 
range plotted in Fig.\ \ref{xueFig4}), the zero-bias conductance is very 
small. As the end group changes, the conductance of the molecular junction 
varies by more than two orders of magnitude, from the maximum of 
45.8 ($\mu S$) for the CN$\Phi$NC molecule to 0.18 ($\mu S$) 
for F$\Phi$F molecule. 

\begin{figure}
\includegraphics[height=3.0in,width=3.0in]{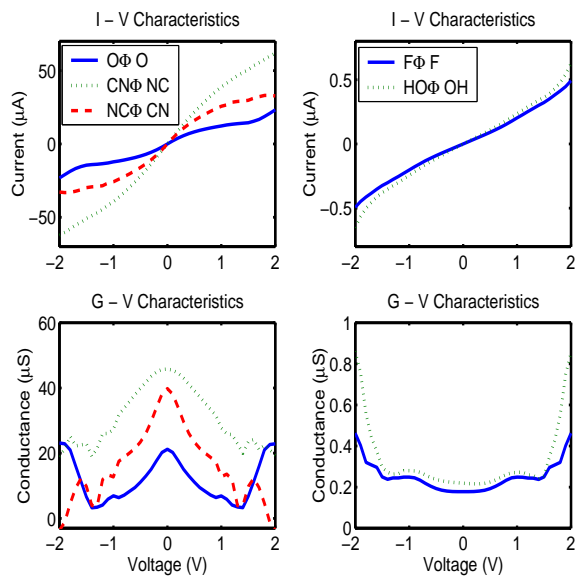}
\caption{\label{xueFig5} Current-voltage (upper figure) and differential 
conductance-voltage (lower figure) characteristics of the five molecules 
studied in this work.} 
\end{figure}

For the device out of equilibrium, the central quantity is the bias-induced 
charge redistribution within the molecular junction and the spatial 
distribution of the voltage drop, which determine the 
current-voltage (I-V), differential conductance-voltage (G-V) characteristics and 
the bias-induced modification of molecular states (the static Stark effect). 
The I-V and G-V characteristics are calculated self-consistently at each bias 
voltage and shown in Fig. \ref{xueFig5}. For the CN$\Phi$NC and NC$\Phi$CN 
molecules, the conductance peaks near zero-bias are due to 
near-resonant-tunneling through the LUMO. For the O$\Phi$O molecule, it is 
due to near-resonant-tunneling through the HOMO. Within the bias range 
considered here, both the F$\Phi$F molecule and the HO$\Phi$OH molecule 
are insulating due to the small tunneling probability through the HOMO-LUMO 
gap, leading to nearly-linear (Ohmic) I-V characteristics from 
$-1(V)$ to $+1(V)$. By examining the transmission characteristics as a function 
of the applied bias (not shown here), we find that the increase of conductance 
towards $\pm 2(V)$ is due to the closer alignment of the metal Fermi-levels 
with the LUMO for the F$\Phi$F molecule and the HOMO for the HO$\Phi$OH 
molecule.  

\begin{figure}
\includegraphics[height=3.0in,width=3.0in]{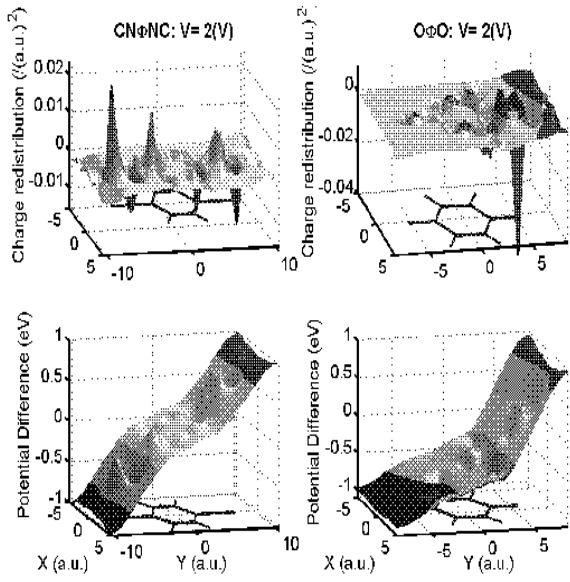}
\caption{\label{xueFig6} (Color online) Charge redistribution and cross-sectional view 
of the voltage drop across the gold-CN$\Phi$NC-gold (left figure) and 
the gold-O$\Phi$O-gold (right figure) junction at bias voltage of $2(V)$. The 
transferred charge is obtained by integrating the difference in the electron density 
at finite and zero biases and plotted as a function of position in the xy-plane. }
\end{figure}

Although the device structures considered here are all symmetric 
with respect to the exchange of the source and drain electrodes, the 
voltage drop across the metal-molecule-metal junction at finite bias voltage 
shows different behavior depending on the potential landscape 
throughout the equilibrium molecular junction (the voltage drop is obtained by 
evaluating the difference between the electrostatic potential at finite and zero 
biases, which obeys the boundary condition of approaching $\pm V/2$ inside the 
electrodes). For the CN$\Phi$NC, NC$\Phi$CN and HO$\Phi$OH molecules 
where there is a narrow potential barrier at the end group region but not in the 
molecule core at equilibrium, the voltage drops (nearly) equally between the 
two metal-molecule interfaces. But for the O$\Phi$O molecule where the potential 
barrier for electron injection is much larger and extends throughout the 
molecule (end group plus molecule core) region at equilibrium, most of the 
voltage drop occurs at the interface between the molecule and the source 
(electron-injecting) electrode, as illustrated in Fig. \ref{xueFig6} 
for bias voltage of $2(V)$. This is readily analyzed using the concept of 
resistivity dipole for nonequilibrium transport systems in the presence of 
current flow,~\cite{Xue031,Landauer} where charge dipoles and 
correspondingly local electric fields develop in the vicinity of scattering 
centers (potential barriers here) to ensure the current continuity 
throughout the conduction system. For the O$\Phi$O molecule with a large 
and thick potential barrier for electron injection, a large electrical field 
is required across the molecule-right electrode interface at positive bias 
voltage, which is sustained by a large electron deficit on 
the molecule side and leads to a large amount of the voltage drop across the 
corresponding interface (similar situation occurs at the molecule-left 
electrode interface at negative bias). In the case of the 
CN$\Phi$NC, NC$\Phi$CN and HO$\Phi$OH molecules, only a relatively 
small electric field is needed for electrons to penetrate a narrow barrier 
at the molecule-source electrode interface, so the electron 
redistributions around the two metal-molecule interfaces have similar 
magnitude and voltage drops nearly equally across the two interfaces. 
For the F$\Phi$F molecule where there is an electrostatic potential well 
in the equilibrium junction, the voltage also drops equally between the 
two metal-molecule interfaces due to the absence of barrier. In addition, 
the difference in the potential response of the molecules to the applied 
bias voltage leads to different behavior in the bias-induced shift of 
molecular levels. For the CN$\Phi$NC, NC$\Phi$CN, HO$\Phi$OH and F$\Phi$F 
molecules, the energy levels of HOMO and LUMO are nearly constant within 
the bias voltage range considered here. But for the O$\Phi$O molecule where 
the voltage drops mostly across the molecule-source electrode interface, 
the shift of the HOMO and LUMO levels follows the Fermi-level of the 
drain (electron-extracting) electrodes. 

\emph{Discussion and Conclusion} Most current experimental works on molecular 
electronics have focused on molecules attached to the gold electrodes through 
sulfur end group,~\cite{Cahen03} due to the convenient thiol-gold self-assembly 
scheme. In this work, we have found theoretically that depending on the 
end groups used, the molecules with identical molecule cores can show 
distinctly different behavior in their response to the perturbation induced by the 
metal-molecule interaction and the applied bias voltage, leading to widely 
different transport characteristics. Since one of the major advantages of molecular 
electronics is the potential to build devices with the desired properties from the 
bottom up, it is important to explore the feasibility of such device engineering 
through molecular design of interfaces. Although we are not aware of any existing 
single-molecule measurements that test the end group effects as reported 
here,~\cite{Frisbie02} such tests should be plausible, in particular using the 
mechanical break junction techniques.~\cite{Weber02}

This work was supported by the DARPA Moletronics program, the DoD-DURINT 
program and the NSF Nanotechnology Initiative. 

\newpage

\end{document}